\documentclass[aps,prl,twocolumn]{revtex4}

\usepackage{amsmath}
\usepackage{amsfonts}
\usepackage{amssymb}
\usepackage{graphicx}
\usepackage{dcolumn}
\usepackage{bm}
\usepackage{hyperref}

\usepackage[tight]{subfigure}

\newcommand{\msec}{\,\mbox{ms}}

\newcommand{\hz}{\,\mbox{Hz}}
\newcommand{\khz}{\,\mbox{kHz}}
\newcommand{\kHz}{\khz}

\newcommand{\nm}{\,\mbox{nm}}
\newcommand{\mcm}{\,\mu\mbox{m}}

\newcommand{\wrho}{\omega_\rho}
\newcommand{\wz}{\omega_z}
\newcommand{\vecr}{\left(\vec{r}\right)}

\newcommand{\ket}[1]{\left| #1 \right\rangle}
\newcommand{\expectation}[1]{\left\langle #1 \right\rangle}

\newcommand{\cre}[2]{\hat{#1}^{\dag}_{#2}}
\newcommand{\ann}[2]{\hat{#1}_{#2}}
\renewcommand{\vec}[1]{\mathbf{#1}}
\newcommand{\abs}[1]{\left|#1\right|}
\newcommand{\half}{\frac{1}{2}}

\newcommand{\fig}[1]{Fig.~\ref{#1}}
\newcommand{\eq}[1]{Eq.~\ref{#1}}

\def\nat{ Nature (London) }

\def\pra{ Phys.\ Rev.\ A }

\def\prl{ Phys.\ Rev.\ Lett.\ }

\def\njp{ New\ J.\ Phys.\ }

\begin{document}
\title{Damping of bulk excitations over an elongated BEC - the role of radial modes }
\author{E. E. Rowen}
\email{eitan.rowen@weizmann.ac.il}
\author{N. Bar-Gill}
\author{R. Pugatch}
\author{N. Davidson}
\affiliation{Weizmann Institute of Science, Rehovot, Israel,
76100}
\date{\today}
\begin{abstract}
We report the measurement of Beliaev damping of bulk excitations
in cigar shaped Bose Einstein condensates of atomic vapor. By
using post selection, excitation line shapes of the total
population are compared with those of the undamped excitations. We
find that the damping  depends on the initial excitation energy of
the decaying quasi particle, as well as on the excitation
momentum. We model the condensate as an infinite cylinder and
calculate the damping rates of the different radial modes. The
derived damping rates are in  good agreement with the
experimentally measured ones. The damping rates strongly depend on
the destructive interference between pathways for damping, due to
the quantum many-body nature of both excitation and damping
products.
\end{abstract}
\maketitle

Ever since Wigner and Weisskopf first calculated the damping rate
of an atom coupled to the vacuum \cite{wignerWeisskopf}, it is
known that coupling to a bath affects the decay  of an otherwise
stable quantum system. The structure of the bath plays a key role
in determining the rate of the damping, and may even cause the
damping to be sub-exponential or super-exponential
\cite{kuritzki:zeno}. Gaseous Bose Einstein condensates (BEC) are
usually well isolated from their surroundings, leading to coherent
evolution. Three body loss, which is the main cause for
decoherence in the ground state, can be on the order of seconds
\cite{cornell:3bodyloss}. However, damping of the elementary
excitations over the condensate is
 much more rapid, but still accessible experimentally.
 The bath, which is coupled to the excitation, is in this case a quasi-continuum of unoccupied excitations.
 These excitations, known as Bogoliubov
quasi particles, are approximate eigen states of the Bosonic
many-body Hamiltonian. Coupling of an excited quasi particle (QP)
to initially unoccupied QP modes gives rise to decoherence via the
Beliaev damping mechanism, in which a QP decays into two new QPs
while fulfilling energy and momentum conservation
\cite{beliaev:beliaev}. Another damping process, that takes place
at a finite temperature, is called Landau damping. This process
involves the annihilation of a thermally excited QP together with
the damped one, and the creation of a more energetic QP, also
conserving momentum and energy \cite{stringari:landau}.

Discrete Beliaev coupling of excitations was measured in
\cite{foot:beliaev,dalibard:quadropole}. Since the spectrum
is discrete, such Beliaev coupling of low energy excitations is
rare, and even at low temperatures, damping is usually governed by
Landau processes
\cite{stringari:landau,walraven:damping,zaremba:landau,pitaevskii:quadropoleInfiniteCylinder,giorgini:MFdamping}.
Damping of bulk excitations is due mainly to Beliaev coupling at
low temperatures \cite{giorgini:MFdamping}. Such damping was
measured \cite{ours:beliaev}, and found to have strong momentum
dependence, in agreement with \cite{beliaev:beliaev}. However, the
dependence of the Beliaev damping on the excitation energy was not
measured.

In this letter we study the effects of the damping on the
excitation line shape in an elongated condensate.  By  comparing
the overall response to Bragg excitations, with the response of
the undamped fraction, the damping rates are quantified as a
function of both momentum and energy of the decaying QP. In
addition to the momentum dependence of the damping rate, we
observe a dependence on the excitation energy. By modelling the
condensate as an infinite cylinder
\cite{njp.tozzo.1367-2630-5-1-354},\cite{stringari:elongated}, we
calculate the damping rates of the different elementary modes, and
find good agreement with the measured line shapes. According to
the model, the spatial dependence of the QP wave functions plays a
key role in the difference between the damping rates. The
different energy dependence at different momentum excitations is a
result of quantum interference due to  many-body effects.

We create a BEC of $N=4\times10^5$ $^{87}$Rb atoms in the internal
state $\ket{F=2,m_F=2}$ in a cylindrically symmetric
Ioffe-Pritchard trap with trapping frequencies of
$\wrho=2\pi\times 350 \hz$ and $\wz=2\pi\times 30\hz$. The
Thomas-Fermi radii of the condensate are $R_{TF}\approx3 \mcm$ and
$Z_{TF}\approx 36 \mcm$, and the  chemical potential is
$\mu/h\approx 4\khz$. We excite the condensate by shining it with
two off resonant laser beams detuned by $0.2 \nm$ from the D2
transition for a period of $2 \msec$ (Bragg excitation). Then we
rapidly shut the magnetic trap off, and image the atomic cloud
after $38 \msec$ time of flight. By varying both the frequency
detuning $\omega$ and the angle $\alpha$ between the two beams, we
can measure the response of the condensate to an excitation with
momentum $\hbar k=2\hbar k_L\sin(\alpha/2)$ in the axial direction
$\hat{z}$, and with energy $\hbar \omega$. Here $k_L=2\pi/780
\nm^{-1}$ is the wave number of the lasers. Since $k_L Z_{TF}\gg
1$, the wave vector of the excitation $k$ is a good quantum
number.

\begin{figure}[ht]
\begin{center}
\includegraphics[width=8cm]{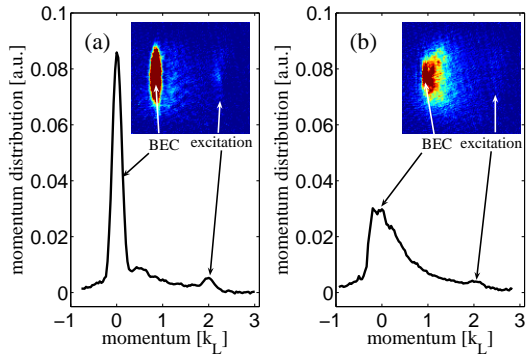}
\end{center}
\caption[Beliaev damping of different radial
modes]{\label{fig:imageshighk} Beliaev damping of condensates
excited with momentum $2\hbar k_L$, and different frequencies:
$14$ kHz (a) and $18$ kHz (b). Insets - absorption images of
excited condensates after $38 \msec$ time of flight. The graphs
show the normalized momentum distribution $p(k)$ along the axial
direction $\hat{z}$, which is a vertical integral of the
absorption images in the insets. The free-particle resonance is at
$15 \kHz$ and the local density resonance for our condensate is
$17.5  \kHz$. In (a) the undamped excitation is larger than that
of (b), while the cloud of Beliaev damping products is larger in
(b). As the Beliaev damping extracts atoms from the condensate,
the condensate peak is greatly suppressed in (b).}
\end{figure}

A unique feature of experiments with BEC is the ability to observe
the damping products in a single image. In the insets of
\fig{fig:imageshighk} we present time of flight images of an
excited condensate. In both images roughly one third of the
condensate atoms were excited to a QP mode with momentum $2\hbar
k_L$, but in  \fig{fig:imageshighk}b, the damping is much larger.
This is evident in the plot of the normalized momentum
distribution in \fig{fig:imageshighk}. For the lower energy
excitation (a), the peak in the momentum distribution at $k=2k_L$
is more pronounced. For the larger energy excitation (b), the
damping products are dominant. The strong reduction of the
condensate fraction along with the tendency of the damping
products toward lower axial momentum are evidence of multiple
collisions.

In order to quantify the damping of the excitations at different
frequencies we employ a post-selection technique
\cite{ours:decoherence}. We measure the response in two ways.
First we measure the average momentum along the $\hat{z}$ axis, of
all atoms in the expanded atomic cloud: $\expectation{k}/2k_L=\int
dk \ k\cdot p(k)/2k_L$, where $p(k)$ is the momentum distribution.
Since momentum along $\hat{z}$ is conserved during damping, this
is a measure of the excitation fraction including damped
excitations and will be referred to as overall response. The
second way to measure response is to count the fraction of atoms
that remain in the excitation mode alone:
$N_{2k_L}/N_{\text{tot}}$, where the occupation $N_{2k_L}$ and the
total number of atoms $N_{\text{tot}}$ are determined by a fit to
the momentum distribution. This is referred to as the undamped
population. In the absence of collisions the two measuring methods
are equivalent. The difference between the two measurements is a
measure of the number of excitations that were damped.

\begin{figure}[t]
    \begin{center}

        {\includegraphics[width=8 cm]{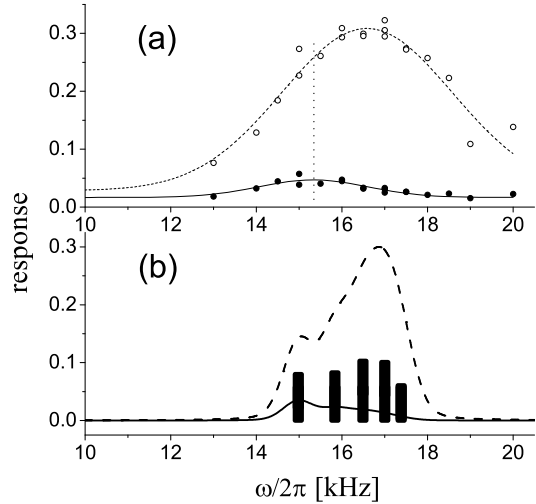}}

    \end{center}
\caption[Beliaev line shape at high k]{(a) Line shape at $k=2k_L$
measured in two ways: ($\circ$) The overall response
$\expectation{k}/2k_L$, and ($\bullet$)  the undamped population
$N_{2k_L}/N_{\text{tot}}$. The dashed and solid lines are gaussian
fits centered at $16.6$ kHz and $15.3$ kHz, with widths of $4$ kHz
and $2.6$ kHz respectively. The resonance in the line shape of the
undamped population is shifted down in energy and is narrower than
that of the overall response. The dotted line marks the center of
the resonance of the undamped population. (b) The corresponding
theoretical curves obtained from our model, for the line shape of
the overall response (dashed line), and that of the undamped
population (solid line). The radial modes leading to the line
shape are marked by bars, positioned at their corresponding
energy, and with a height proportional to their overlap with the
condensate. }\label{fig:highk}
\end{figure}

We repeat this measurement for different values of $\omega$. In
\fig{fig:highk}a we compare the obtained line shape for
excitations with momentum $2\hbar k_L$. Empty circles are the
measurements of the overall response, while the filled circles are
the undamped results of the same images. The overall response
 displays a peak at $16.6\kHz$, less than predicted by the local density approximation
\cite{zarembaDynStructFacPRA.61.063608}. This is due to the
relatively large excitation fraction, which shifts the resonance
towards the free particle value \cite{ours:decoherence}. The width
of the resonance is due mainly to the inhomogeneous density of the
condensate. The response of the undamped population (filled
circles) peaks at $15.3\kHz$, and is 5 times smaller than that of
the total population, due to Beliaev damping. There is a clear
shift down in the resonance   of the undamped population, implying
that the damping rate is larger for the more energetic
excitations. This can be understood intuitively in a local density
approximation: more energetic excitations are in spatial regions
of larger density, leading to a higher damping rate.

\begin{figure}[ht]

    \begin{center}

        {\includegraphics[width=8 cm]{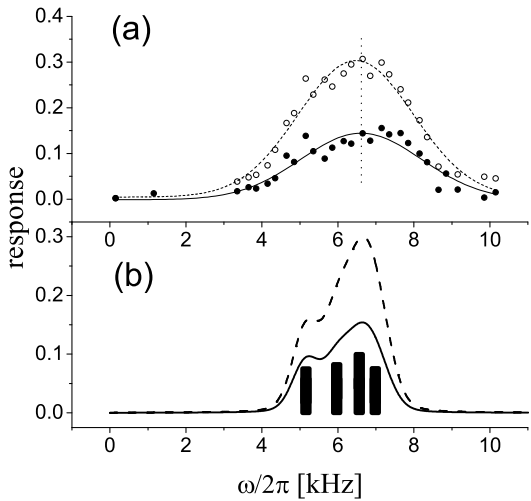}}

    \end{center}
\caption[Beliaev line shape at low k]{ (a) Line shape at
$k=1.1k_L$ measured both ways: ($\circ$) The overall response
$\expectation{k}/1.1k_L$, and ($\bullet$) the undamped population
$N_{1.1k_L}/N_{\text{tot}}$. All circles are averages over 4 data points.
The dashed and solid lines are gaussian fits giving a central
frequency of $6.5\khz$ for the overall response and $6.6\khz$ for
the undamped population.  The dotted line marks the center of the
line shape of the undamped population. There is no downward shift
of the resonance, and no narrowing as are present in
\fig{fig:highk}. (b) The corresponding theoretical curves obtained
from our model, for the line shape of the overall response (dashed
line), and the  undamped population (solid line). The bars mark
the frequency and overlap with the condensate of the relevant
radial modes. }\label{fig:lowk}
\end{figure}
 In Fig.~\ref{fig:lowk}a we present the response of the condensate to
excitations with smaller momentum - $1.1\hbar k_L$. A reduction in
the damping as compared to the case of $2\hbar k_L$ is evident.
This is a result of the quantum interference due to the collective
nature of the low momentum excitations and therefore is present in
a homogeneous BEC as well \cite{ours:beliaev}. A gaussian fit to
the overall  response is centered at $6.48\pm 0.04$ kHz. The fit
to the undamped line shape has a peak at $6.63\pm 0.07$ kHz.
Contrary to the $2k_L$ case, the line shape of the undamped
population is not shifted down in energy, and the naive linear
dependence of the damping rate on the local density fails.

To theoretically account for these effects we include spatial
dependence. We exploit the cigar shape of the condensate to
neglect the $\hat{z}$ dependence of the ground state, thus
reducing the problem to 1D in the radial direction
\cite{njp.tozzo.1367-2630-5-1-354}. A similar treatment is
performed in \cite{pitaevskii:quadropoleInfiniteCylinder} to
describe Landau damping of quadrupole oscillations of an elongated
BEC. The ground state $\psi_0(\rho)$ of the Gross-Pitaevskii
Hamiltonian $H_{GP}=-\nabla^2/2M+V(\rho)+gN/L\abs{\psi_0}^2$ is
calculated by imaginary time evolution, for a radially dependent
potential $V(\rho)=\half M\omega_\rho^2 \rho^2$ and our
experimental parameters. Excitations over the condensate are
obtained by solving the Bogoliubov equations:
\begin{eqnarray} \hbar \omega_\vec{\nu} u_{\vec{\nu}}\vecr =
\left[H_{GP}+g\frac{N}{L}\abs{\psi_0}^2\right] u_{\vec{\nu}}
+g\frac{N}{L}\psi_0^2 v_{\vec{\nu}} \label{eq:bogu}\\
 -\hbar \omega_\vec{\nu}
v_{\vec{\nu}}\vecr =
\left[H_{GP}+g\frac{N}{L}\abs{\psi_0}^2\right]v_{\vec{\nu}}
+g\frac{N}{L}{\psi_0^*}^2 u_{\vec{\nu}}.\label{eq:bogv}
\end{eqnarray}
 The wave functions $u_{\vec{\nu}}\vecr$ of
Eqs.~\ref{eq:bogu},\ref{eq:bogv} are decoupled to
$u_{\vec{\nu}}\vecr=u_{n,m,k}(\rho)e^{im\phi}e^{ikz},$ and so are
$v_{\vec{\nu}}\vecr$. The set of quantum numbers ${\vec{\nu}}$ are
in this case: $k$ - the momentum along $\hat{z}$, $m$ - the
vorticity around $\hat{z}$, and $n$ - the number of radial nodes
of the wave functions $u(\rho)$ and $v(\rho)$. The energy of the
radial modes $\hbar\omega_{k,m,n}$ increases with $n$ as well as
$\abs{k}$ and $\abs{m}$. There is a conservation law for $k$ and
$m$ ($k$ can only decay into $q,k-q$, and $m$ into $m_1,m_2$
fulfilling $m_1+m_2=m$), but the quantum number $n$ is not
conserved. We consider the damping of excitations with $m=0$ as
the Bragg pulse carries no angular momentum along $\hat{z}$.
Still, all $m$ values need to be considered as damping products.

The Beliaev coupling is part of the next order expansion of the
many-body Hamiltonian:
$H_B=\sum_{{\vec{\nu}},{\vec{\nu}}_1,{\vec{\nu}}_2}A^{\vec{\nu}}_{{\vec{\nu}}_1;{\vec{\nu}}_2}\cre{b}{{\vec{\nu}}_1}\cre{b}{{\vec{\nu}}_2}\ann{b}{{\vec{\nu}}}+\text{H.C.}$
The three QP overlap
\begin{eqnarray}\label{eq:Anu}
A^{\vec{\nu}}_{{\vec{\nu}}_1;{\vec{\nu}}_2}=2g\sqrt{N_0}\int
d\vec{r} \left(u_{\vec{\nu}}
u_{{\vec{\nu}}_1} u_{{\vec{\nu}}_2}+ u_{\vec{\nu}} u_{{\vec{\nu}}_1} v_{{\vec{\nu}}_2}\right.\nonumber\\
 \left.+ u_{\vec{\nu}} v_{{\vec{\nu}}_1} u_{{\vec{\nu}}_2}+ v_{\vec{\nu}} u_{{\vec{\nu}}_1} v_{{\vec{\nu}}_2}+
v_{\vec{\nu}} v_{{\vec{\nu}}_1} u_{{\vec{\nu}}_2}+ v_{\vec{\nu}}
v_{{\vec{\nu}}_1} v_{{\vec{\nu}}_2}\right)
\end{eqnarray}
 involves interference of six different quantum pathways.

\begin{figure}[t]
    \begin{center}
    \includegraphics[width=8 cm]{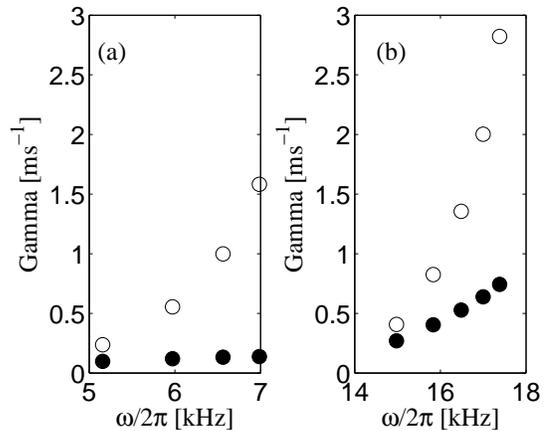}
    \end{center}
\caption[model damping rates]{\label{fig:gamma} Damping rates
versus the energy of the different radial modes for the first few
radial modes with $k=1.1k_L$ (a), and $k=2k_L$ (b). ($\bullet$) -
the calculated damping rate. ($\circ$) - the damping rate obtained
by taking only the terms involving $u_{\vec{\nu}}
u_{{\vec{\nu}}_1} u_{{\vec{\nu}}_2}$ in \eq{eq:Anu}.}
\end{figure}

Using first order time dependent perturbation theory we calculate
the damping as a function of time. Since the damping is nearly
linear, we can extract damping rates
\begin{equation}
\Gamma_{\vec{\nu}}=\frac{1}{t}\frac{4}{\hbar^2}\sum_{m,n_1,n_2}\int
dq \abs{A_{\vec{\nu_1};\vec{\nu_2}}^{\vec{\nu}}}^2
\frac{\sin^2\left(\Delta\omega t/2\right)} {\Delta\omega^2},
\end{equation}
from the different initially excited radial modes
$\vec{\nu}=(n,0,k)$, to all possible pairs of modes
$\vec{\nu_1}=(n_1,m,q)$ and $\vec{\nu_2}=(n_2,-m,k-q)$, with an
energy mismatch of
$\Delta\omega=\omega_{\vec{\nu}}-\omega_{\vec{\nu_1}}-\omega_{\vec{\nu_2}}$
 These rates are plotted in \fig{fig:gamma} versus the corresponding
eigen-frequencies $\omega_{n,0,k}$.

The radial mode dependence of the damping rate is different for
the different initial momenta. The damping rate from radial modes
with momentum $k=2k_L$ increases with $n$. In this regime,
$\hbar\omega\gg\mu$, excitations are nearly single-particle in
nature and $v\vecr\rightarrow 0$, for both the decaying
excitation, and most of the damping products. Therefore
 the damping is mostly governed by the term involving only
$u_{\vec{\nu}} u_{{\vec{\nu}}_1} u_{{\vec{\nu}}_2}$. This can be
seen by the fact that the damping rates in \fig{fig:gamma}b
(filled circles), follow the term with only $u$'s (empty circles)
\cite{footnote}. The increase of both rates with radial mode
energy is a result of the number of available energy-conserving
pairs of damping products. As the energy of the decaying QP is
increased, more pairs are available, and the damping increases.
Our calculations show that the $n=4$ mode has twice as many
possible damping products as the mode $n=0$, in agreement with the
damping rate which is approximately doubled.

Damping of modes with momentum $1.1\hbar k_L$ is different. First,
the energy spacing between modes is comparable to the $2k_L$ case,
but the energies are smaller. This is manifested in a
\emph{larger} relative increase of available modes with radial
mode. In fact, the $n=3$ mode has 5 times more possible damping
target modes than the $n=0$ mode. Alone, this would increase the
damping from the higher radial modes as happens with the empty
circles in \fig{fig:gamma}a. This effect is compensated by
interference between the different terms in \eq{eq:Anu}. Nearly
all the damping products  have $\hbar\omega\lesssim\mu$, and
therefore, as the radial number increases, the wave functions
overlap denser regions of the condensate and $v\vecr$ becomes more
significant. Together, the increase due to phase space, and the
reduction due to quantum interference between the terms in
\eq{eq:Anu} cancel, and the damping is similar from all radial
modes for $k=1.1k_L$  as seen in filled circles of
\fig{fig:gamma}.

 For even smaller momentum excitations, our model predicts non-exponential damping.
  However, the damping is so slow, that the infinite cylindrical approximation is no longer valid for our trap parameters.

The line shapes of the overall response, and the undamped
population obtained by the model are presented below the data  in
Figs.~\ref{fig:highk}b,~\ref{fig:lowk}b. The bars are positioned
at the eigen-frequencies of the radial modes with the
corresponding $k$ (for $m=0$). The height of the bars is
proportional to the overlap with the condensate, that determines
the response to the Bragg pulse
\cite{njp.tozzo.1367-2630-5-1-354}. The dashed lines are the
expected overall response to the  Bragg pulse, obtained by
summation over the responses of the different radial modes
including Fourier broadening of each mode, as in
\cite{njp.tozzo.1367-2630-5-1-354}. The solid line is obtained in
a similar manner after multiplying each response function by the
corresponding damping $e^{-\Gamma_{n,0,k}t_{\text{eff}}}$. We
further convolve the line shape with a gaussian of width $300\hz$
to account for residual sloshing in the trap, that changes the
effective detuning \cite{ours:multibranch}. The suitable effective
damping time, $t_{\text{eff}}=5 \msec$, is $3 \msec$ longer than
the Bragg pulse. This may be due to collisions during the
expansion of the condensate and enhancement of the damping due to
thermal population of the damping products.

 Gaussian fits to the
calculated line shapes give resonances at $6.3 \khz$ for both
overall response and undamped population for the $1.1k_L$
excitations. For the $2k_L$ excitations, the model gives a line
shape centered at $16.5 \khz$ for the overall response and at
$15.6 \khz$ for the undamped population in good agreement with the
measured values. The model displays the same narrowing and
downward shift in the line shape of the undamped population at
high momentum, indicating the larger damping rate from more
energetic excitations. Both model and experiment also show a
uniform, much smaller damping of the small momentum excitations.
The lowest radial mode is more separated from the others, and
distinguishable from the rest both in  the model and in the
experimental data.

In conclusion, we use a post selection technique to measure the
Beliaev damping of QPs in an elongated condensate. We find a
dependence of the damping rate on both  the excitation energy and
the momentum of the decayed QP. We measure a shift downward in the
excitation energy of the undamped excitation for high momentum,
while for small momentum we measure no such shift. Modelling our
system as infinite along the axial direction, we obtain
quantitative agreement between theory and experiment. The overall
damping is suppressed for low momentum excitations as a result of
interfering quantum pathways within each damping channel.

 The
inhomogeneous treatment includes a quantization of the bath in the
radial direction. Since our excitation energies are above
$\hbar\omega_\rho$, we are only indirectly sensitive to this
quantization. Beliaev damping becomes even more intriguing when
the radial trapping frequency is increased and the system becomes
one dimensional. In this regime it could be possible to observe
effects such as quantum Zeno \cite{kuritzki:zeno}, non exponential
damping and even the complete inhibition of damping due to the
convex excitation line shape.

Studying damping in our system has the advantage of imaging which
modes of the continuum are excited upon damping. This can serve as
a spectroscopic tool to probe the QP spectrum of the damping
products \cite{ours:collisions}. One can also differentiate
between Beliaev damping and Landau damping according to the
resulting momentum distribution of damping products.

 This work was
supported by the DIP  and Minerva foundations.


\begin{thebibliography}{19}
\bibitem[{foo()}]{footnote}
\bibinfo{note}{The discrepancy between the filled and empty circles is
due to destructive interference in damping events involving
damping products with non-negligible $v\vecr$.}
\end{thebibliography}
\end{document}